\documentclass[fleqn,twoside]{article}
\usepackage{espcrc2}

\usepackage{graphicx}
\usepackage{amssymb}

\title{Review on Neutrino Telescopes}

\author{T. Montaruli\address[UW]{Physics Department, 
        University of Wisconsin-Madison, 
        WI-53706, USA}%
        \thanks{on leave of absence from University of Bari, Physics Department and INFN, I-70126.}
   }

\begin{document}

\begin{abstract}
I will discuss the motivations for Neutrino Astronomy and its prospects given the current experimental scenario, which is the main focus of this paper.
I will also go through the first results of the IceCube detector deep in the ice and of the ANTARES undersea telescope underlying complementary aspects,
common and different challenges. It is an exciting time for this science since the first completed undersea detector is successfully taking data and the first cubic kilometer detector is going to be shortly more than half-way from its completion in Antarctica.
 
\vspace{1pc}
\end{abstract}

\maketitle

\section{NEUTRINO ASTRONOMY: WHY?}

Answers to this question may sound rhetorical, nonetheless examples can be found in astronomy history that show that whenever new instruments are pointed
to the sky unexpected discoveries are possible. So the discovery of some astrophysical neutrino source not seen with other messengers such as 
photons or protons is possible. In fact photons or nucleons may not be able to escape outside their sources or they may interact with the radio, infra-red and microwave (CMB) backgrounds during their propagation to us. Neutrinos are good messengers due to their weakly interacting properties and due to the fact that, being neutral, they allow us to point back to their sources. Hence they let us access tens of Gpc regions of the sky, while photons with E $\gtrsim 10$ TeV and protons (E $\gtrsim 10^{19}$ eV)  observations are limited to $\sim 10-100$ Mpc depending on their energy.  On the other hand, the weakly interacting properties 
of neutrinos make their detection more challenging and in fact Neutrino Telescopes (NTs) have been operating now about a decade with no positive observation yet. 
Given the above caveats, I will show here at what level we may expect a positive observation given expectations derived from photon observations.

It is commonly believed that about 10\% of the energy emitted in galactic SN explosions at an approximate rate of 3 per century can 
provide the power needed to account for the observed cosmic rays up to the knee region. The knee seems to be Z dependent, so that the composition
becomes heavy above $\sim 4$ PeV. Since the cosmic ray (CR) spectrum extends to regions well above $10^{17}$ eV it is not possible that these 
Ultra-High Energy Cosmic Rays (UHECRs) are of galactic origin but we have at today no firm explanation on what are the sources of the extra-galactic component nor we have determined their composition. Most of the experiments indicate it is lighter than in the region right above the knee and there is common agreement between HiReS and Pierre Auger Observatory (PAO) on the observation of the GZK cut-off due to interactions of protons with the cosmic microwave background \cite{GZK}. PAO used 27 highest energy events (E$>57$ EeV) and found that 20 are at less than $3.2^o$ from Active Galactic Nuclei (AGN) in their catalogue at a distance $< 71$ Mpc while 5.6 are expected on average. Nonetheless, this is not a conclusive observation that AGN are the sources of cosmic rays. The  isotropic distribution of events is incompatible at 1\% level \cite{PAOanistropies}. On the other hand, HiRes does not
confirm the result \cite{hires_anisotropy}. 

\begin{figure}[htb]
\vspace{9pt}
\includegraphics[width=20pc,height=15pc]{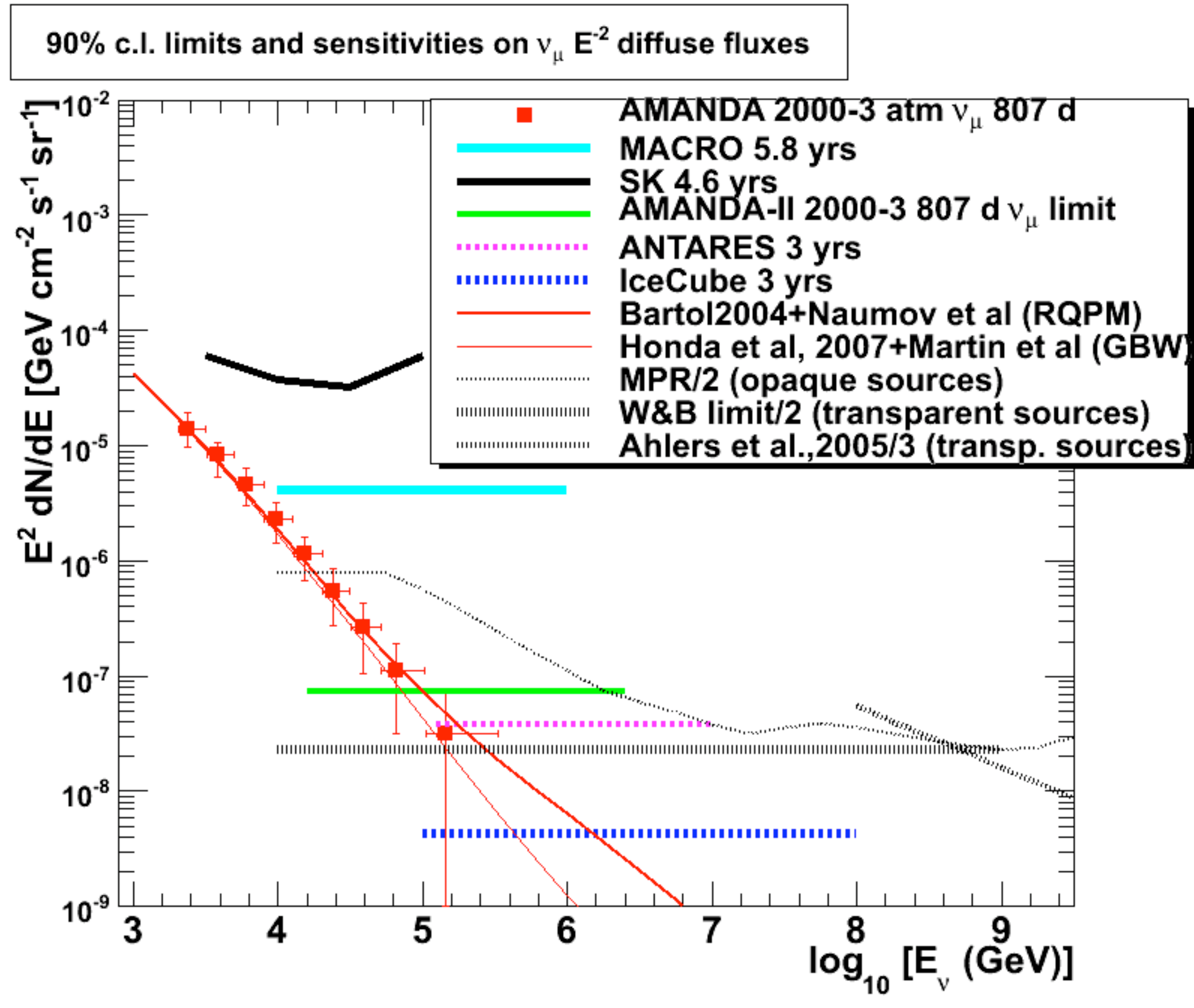}
\caption{Horizontal and dashed lines are upper limits and sensitivities, respectively, to an $E^{-2}$ $\nu_{\mu}+\bar{\nu}_{\mu}$ flux vs $\nu$ energy. From top to bottom: SK \protect\cite{sk_diffuse}, MACRO \protect\cite{macro_diffuse}, AMANDA-II \protect\cite{amanda_diffuse}, and predicted sensitivities for 3 yrs of data taking of ANTARES \protect\cite{antares_diffuse} and IceCube \protect\cite{performance} (bottom dashed line). The W\&B \protect\cite{WB} and \protect\cite{MPR} upper bounds for respectively optically thin and thick sources and the flux in \protect\cite{ahlers} are shown as dotted black lines. 
The measured atmospheric muon neutrino spectrum by AMANDA-II is shown (red squares) \protect\cite{amanda_atmnu} with predicted fluxes for the conventional+prompt neutrinos (upper curve Bartol flux \protect\cite{bartol} + Naumov RQPM \protect\cite{naumov}, lower curve HKKM \protect\cite{HKKM} + Martin et al pQCD model \protect\cite{martin}). }
\label{fig1}
\end{figure}

Upper limits on diffuse $E^{-2}$ fluxes exist from NTs covering the range from about 10 TeV to some PeV for muon neutrinos as well as for neutrino induced cascades of all flavors. Fig.~\ref{fig1} shows the experimental results for muon neutrinos as well as the unfolded atmospheric neutrino
spectrum measured by AMANDA-II. 
Neutrinos could offer a source of further information about the UHECRs in the region of the GZK cut-off. Protons with energies $\gtrsim 10^{19}$ eV interact with the CMB and infrared backgrounds producing pions that decay into neutrinos, known as cosmogenic neutrinos. On the other hand, if the UHE cosmic rays (UHECRs) consist of heavy or intermediate mass nuclei rather than protons, they would generate neutrinos through photodisintegration followed by pion production through nucleon-photon scattering and resulting fluxes would be lower than in the pure proton assumption. Cosmogenic neutrinos are often considered as a guaranteed UHE flux but predicted fluxes vary over orders of magnitudes due to the degree of freedom in defining the transition region between the galactic fading component and the onset of the extragalactic one. They depend on UHECR composition and spectra but these measurements are affected by large systematics \cite{GZK}. Neutrino detection above $10^{17}$ eV may be challenging for existing detectors and new techniques may be needed to firmly establish their existence and to extract useful astrophysical information on source injection spectra, maximum energy for acceleration in these sources, source evolution and composition. 

\begin{figure}[hbt]
\includegraphics[width=20pc,height=14pc]{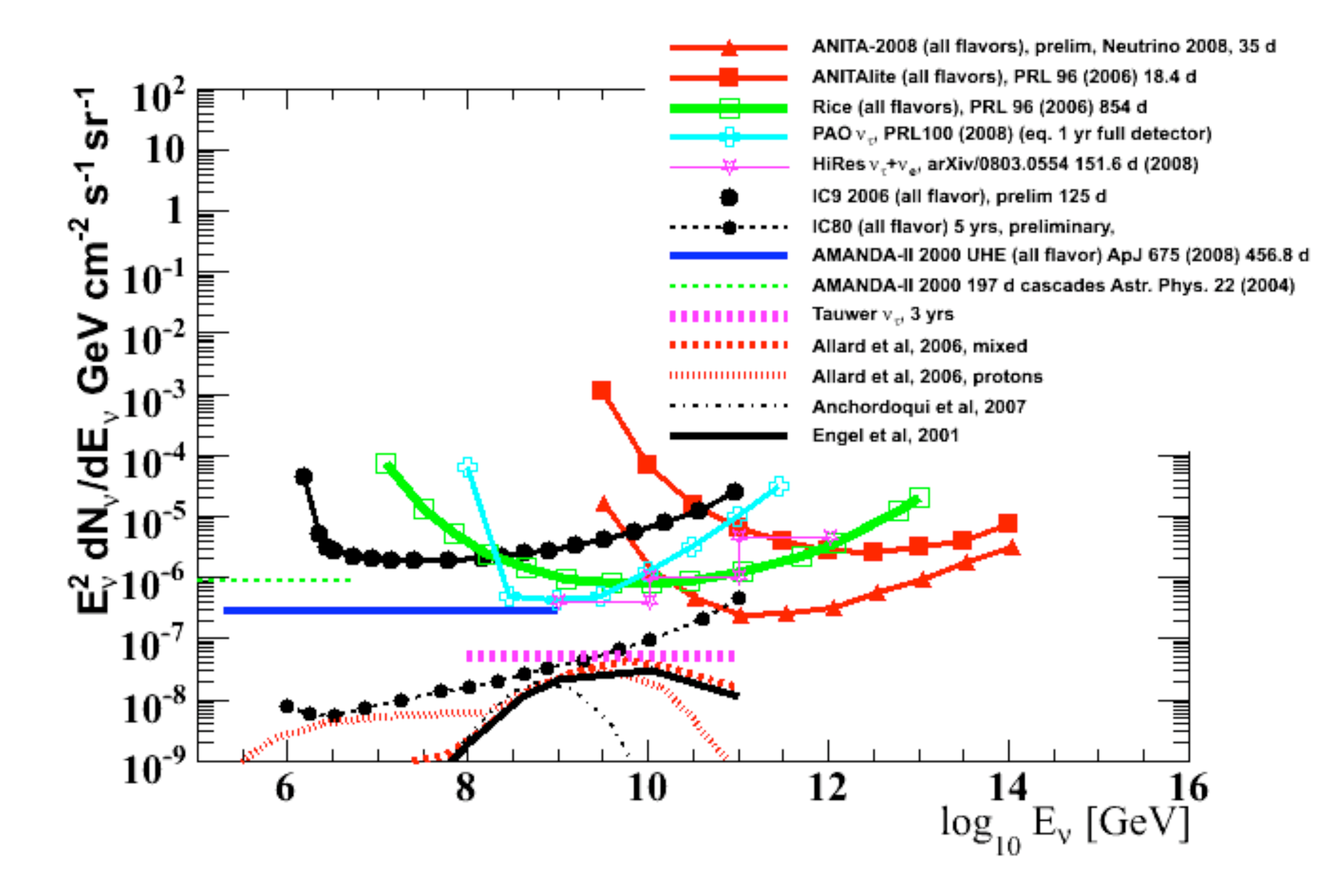}
%\vspace{-9pt}
\caption{90\% c.l. upper limits differential in energy on $\nu_{e}+\nu_{\mu}+\nu_{\tau}$ ($1:1:1$ oscillation assumption) for:  ANITA prototype  (red line-full squares) \protect\cite{anita} and for the Dec 06-Jan 07 flight (red line-full triangles) \protect\cite{gorham}; Rice \protect\cite{rice} (scaled from 95\% c.l. to 90\% for 0 selected events - green open squares) using the radio technique, 
IC9 upper limit (solid black line with full circles),  preliminary estimate for IC80 in 5 yrs (dashed line with full circles) \protect\cite{eheicrc07}.
PAO limit is on the Earth-skimming $\nu_{\tau}$ flux (cyan crosses) \protect\cite{PAO} and HiRes limit is on $\nu_e + \nu_{\tau}$ (pink stars)\protect\cite{hires}. AMANDA UHE (solid horizontal blue line) \protect\cite{amandauhe} and AMANDA cascade limits \protect\cite{cascade}, Tauwer foreseen sensitivity for $\nu_{\tau}$ \protect\cite{tauwer} are for $E^{-2}$ fluxes. Models are from \protect\cite{allard} (pure proton and mixed composition) and \protect\cite{engel01}. }
\label{fig2}
\end{figure}

A well known bound to the extra-galactic diffuse flux of neutrinos was proposed in \cite{WB}. Neutrino fluxes above this bound can be obtained
not only assuming optically thick sources \cite{MPR}, but also optically thin ones and normalizing the extra-galactic UHECR component at $\sim 10^{17}$ eV to the measured spectra by HiReS or AGASA \cite{ahlers}, rather than assuming an $E^{-2}$ proton spectrum normalized at $10^{19}$ eV.
In Fig.~\ref{fig1} we show the upper limits and sensitivities to an $E^{-2}$ flux of muon neutrinos for NTs and compare it to the W\&B bound for optically thin sources \cite{WB} and Manneheim, Protheroe and Rachen one for completely opaque sources \cite{MPR},  as well the all flavor ``low cross-over" scenario flux in \cite{ahlers}. The W\&B flux, after accounting for oscillations, corresponds to about 50 muon neutrino events in the full IceCube per year. In Fig.~\ref{fig2} we show the experimental bounds on cosmogenic neutrino fluxes for various experiments. No correction is applied to upper limits that are sensitive only to 1 or 2 flavors respect to those sensitive to 3 flavors since exposures for different flavors differ.

\section{DETECTORS AND SELECTED RESULTS}

\begin{figure}[htb]
\vspace{9pt}
\includegraphics[width=20pc,height=15pc]{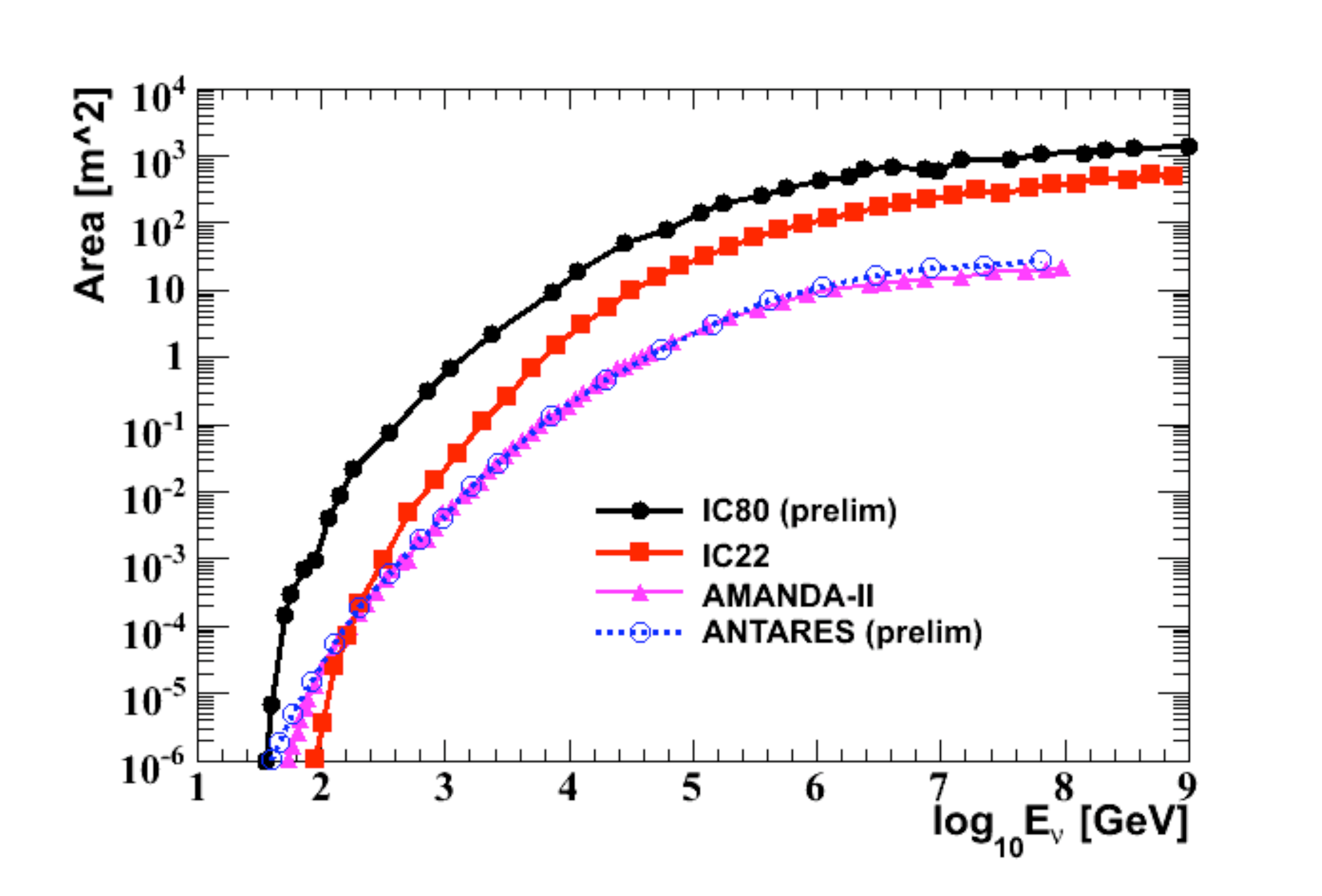}
\caption{ $\nu_{\mu}+\bar{\nu}_{\mu}$ effective area vs neutrino energy for the lower hemisphere for ANTARES, AMANDA-II, IC22 and IC80. 
For AMANDA-II and IC22 these areas correspond to optimized point-like source data analyses. IC80 area is obtained with the same cuts studied for IC22 and 
ANTARES one is obtained for point source analysis cuts studied by simulation. The AMANDA-II and ANTARES lines stop at $10^{8}$ GeV because this is the upper limit of the simulations used for obtaining them.}
\label{fig3}
\end{figure}

\begin{figure}[htb]
%\vspace{9pt}
\includegraphics[width=20pc,height=15pc]{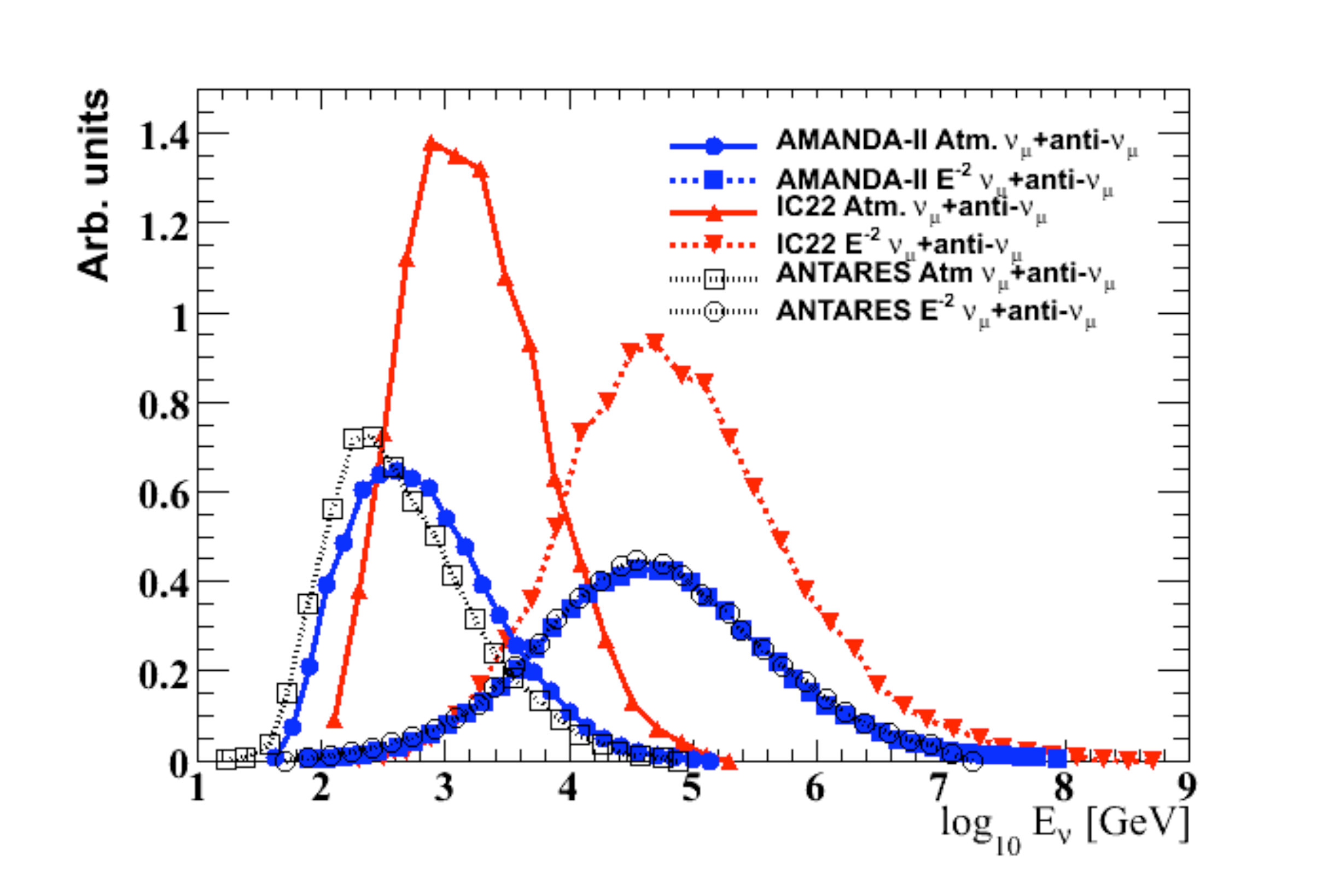}
\caption{Response curves (arbitrary units) for ANTARES, AMANDA and IC22 for atmospheric neutrinos \protect\cite{bartol} and for an $E^{-2}$ spectrum (same selections than in Fig.~\protect\ref{fig3}). The smaller spacing and larger photocathode area density of ANTARES and AMANDA makes their detection threshold lower than for IC22. IceCube is capable of measuring events of much higher energy than AMANDA or ANTARES and will have better performance at low energies than what shown here with the addition of DeepCore. }
\label{fig4}
\end{figure}

NTs use a 3D array of photomultipliers (PMTs) to detect the Cherenkov light produced by charged particles with velocity in the radiator (ice or water) larger than that of light. The spacing between strings holding sensors is determined by the properties of the radiator, by costs and by the energy region of interest. Experiments that focus on extending observations well above 10 PeV require cubic-kilometer dimensions and hence cost is the limiting factor that determines the spacing of optical modules (OMs), pressure resistant glass spheres that house PMTs. The photocathode area density determines the reconstruction performance of these detectors that is energy dependent. 
Most of the information that qualify detectors is contained in the effective area for neutrinos (see Fig.~\ref{fig3}), a parameter that for muon neutrinos is given by the probability of the charge current interaction in the target medium, the selection efficiency and the muon range. This parameter convoluted to a neutrino flux will return the observed rate of events from that flux. Since the neutrino cross section is small ($\sim 10^{-34}$ cm$^2$ at $E_{\nu} \sim 30$ TeV, increasing almost linearly with energy), the equivalent size of these detectors translates into very small areas compared to their geometrical size. Because this area is strongly dependent on energy and spans several decades in energy, different neutrino fluxes will produce differential distributions of events vs energy (response curves) that depend on the flux slope. In Fig.~\ref{fig4} we show the response curves for ANTARES, AMANDA-II and IC22 for the atmospheric neutrino spectrum \cite{bartol} and for a harder spectrum $E^{-2}$ close to what is expected from $1^{st}$ order Fermi acceleration mechanisms. For IC22 90\% of the upgoing muons are in the range 3 TeV-3 PeV for $E^{-2}$ and between 250 GeV - 16 TeV for atmospheric neutrinos.
Since the peak of the response curve moves to higher energies the harder the spectrum, correspondingly the area of interest is higher for most of the signal events compared to atmospheric neutrinos, indicating that this technique is optimized for high energies due to the rise of cross sections and of the muon range with neutrino energy. Nonetheless, atmospheric neutrino events dominate since incident fluxes in the detector have much higher normalization.  The ANTARES detector is slightly more sensitive to lower energies for atmospheric neutrinos than AMANDA-II due to the fact that in ANTARES there are 3 close-by optical modules at each storey vertically separated along a string by 14.5 m while in AMANDA there is only one with a vertical separation of about 10 m. String separation is smaller in AMANDA by on average 
about 15-20 m but though in ice the absorption length is larger than in sea water, the scattering length is much smaller. 

AMANDA-II, IceCube precursor, at a depth between about 1.5 to 2 km, is made of 19 strings at an average distance of 50 m forming a cylinder of 200 m and holding 677 OMs containing 20 cm-diameter PMTs. The newer generation detector, IceCube, also located at the South Pole, uses the digital technology to digitize the signals from the 25 cm in diameter PMTs with 300 MHz custom chips and 40 MHz Flash ADCs. The IceCube Observatory is composed of a deep detector made of instrumented strings between about 1.5 and 2.5 km and a surface array called IceTop. IceCube strings are 1 km long, hence IceCube  profits compared to AMANDA of the higher transparency of deep ice and of the larger horizontal acceptance relevant for neutrinos of energies $\gtrsim 1$ PeV since earth shadowing effects become more severe for vertical directions compared to horizontal ones. Each string holds 60 digital OMs (DOMs) separated vertically by about 17 m. Strings are about 125 m apart. The configuration taking now data is comprised of 40 strings (IC40) and 40 IceTop stations corresponding to each string of two frozen water tanks. Each tank contains 2 DOMs that can measure the light emitted by the electromagnetic component of atmospheric showers and the combination of the 2 detectors can measure also the muons that penetrate deep in the ice. IceCube construction will be concluded when 80 strings will be installed at a rate of about 16 per season up to the 2010-11 season. Additional 6 strings, called DeepCore, are planned to be installed with 50 out of 60 of high quantum efficiency sensors deployed with vertical spacing of 7m in the clear ice below 2100 m. The region of dense string spacing will consist of a total of 13 strings covering an area of roughly 250 m in diameter. The aim of this densely instrumented array is to lower the threshold of IceCube for enhancing sensitivity in the region below 1 TeV, interesting for dark matter studies, neutrino oscillations or galactic sources with steep spectra or cut-off at about a few TeV. For these studies DeepCore will increase IceCube effective area by about an order of magnitude at 100 GeV.The large instrumented volume of IceCube around DeepCore will offer a muon veto that will allow to single out starting muons in the detector due to low energy neutrino interactions in the fiducial volume.
Another extension under consideration is to enhance the effective area of IceCube for energies $>100$ TeV. A modified arrangement of the outer 12 strings can lead to an increase in effective area between 15-40\% for well reconstructed tracks depending on their distance from the rest of the array. This could be advantageous for the detection of cosmogenic neutrinos (see Fig.~\ref{fig2}).
Recent results of IceCube are presented in these proceedings in Ref.~\cite{berghaus}.

ANTARES, completed in May 2008, has been presented in Ref.~\cite{margiotta} at this conference. The status of R\&D studies towards the construction of a cubic-kilometer detector in the Mediterranean have been presented in Ref.\cite{km3,NEMO}.
ANTARES is made of 12 lines held taught by buoys and anchored at the sea floor connected to the JB that distributes power and data from/to shore. The instrumented part of the line starts at 100 m above sea level so that Cherenkov light can be seen also from this region. Lines are separated by 60-75 m from each other and each of the lines holds 25 floors called storeys. Each storey has 3 10-inch PMTs looking downward at $45^{\circ}$ from the vertical housed inside pressure resistant glass spheres made of two halves closed by applying an under-pressure of 200-300 mbar. Storeys also include titanium containers housing the frontend electronics with a pair of ASIC chips per PMT used for signal processing and digitization that provide the time stamp and amplitude of the PMT signal. Each of the OMs contains a pulsed LED for calibration of the relative variations of PMT transit time and a system of LED and laser Optical Beacons allows the relative time calibration of different OMs. An internal clock system distributes from shore the 20 MHz clock signal, that is synchronized by GPS to the Universal Time with a precision of $\sim 100$ ns. Time calibrations allow a precision at the level of 0.5 ns and a positioning system that includes tiltmeters and compasses giving the orientation of storeys and an acoustic triangulation system of hydrophones and transceivers provide the relative sensor position and the line shape reconstruction. 

\subsection{Results on point-like source searches}

\begin{figure}[htb]
%\vspace{9pt}
\includegraphics[width=19pc,height=18pc]{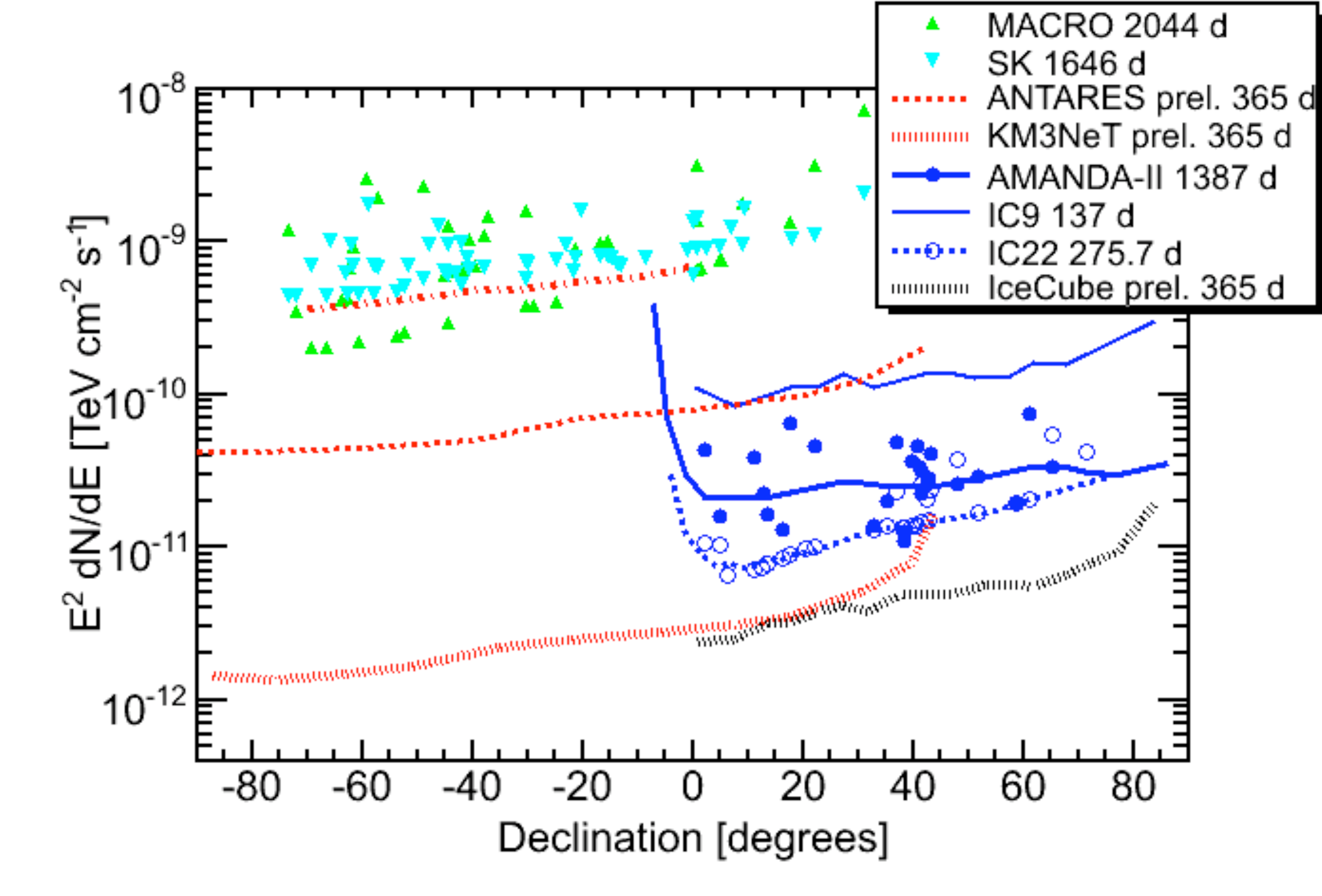}
\caption{Results for point-like source searches for $E^{-2} \nu_{\mu}$ fluxes  vs declination. On the right side from top to bottom: IC9 sensitivity for 137 d \protect\cite{taup07}, AMANDA-II sensitivity and upper limits for specific sources (circles) \protect\cite{amanda7}, sensitivity for 22 strings. For the other hemisphere: sensitivity of 5 lines of ANTARES after 140 d. Data for these configurations have been unblinded while the sensitivity for IC80 and for a reference detector in Ref.~\protect\cite{km3} is from simulation and not optimized. Predicted results  
for 1 yr of the full ANTARES configuration (dashed line).  Upper limits for catalogues of selected  sources are shown for Super-Kamiokande 
(full triangles) \protect\cite{sk} and MACRO (empty triangles) \protect\cite{macro}.}
\label{fig5}
\end{figure}

\begin{figure}[htb]
\vspace{9pt}
\includegraphics[width=20pc,height=15pc]{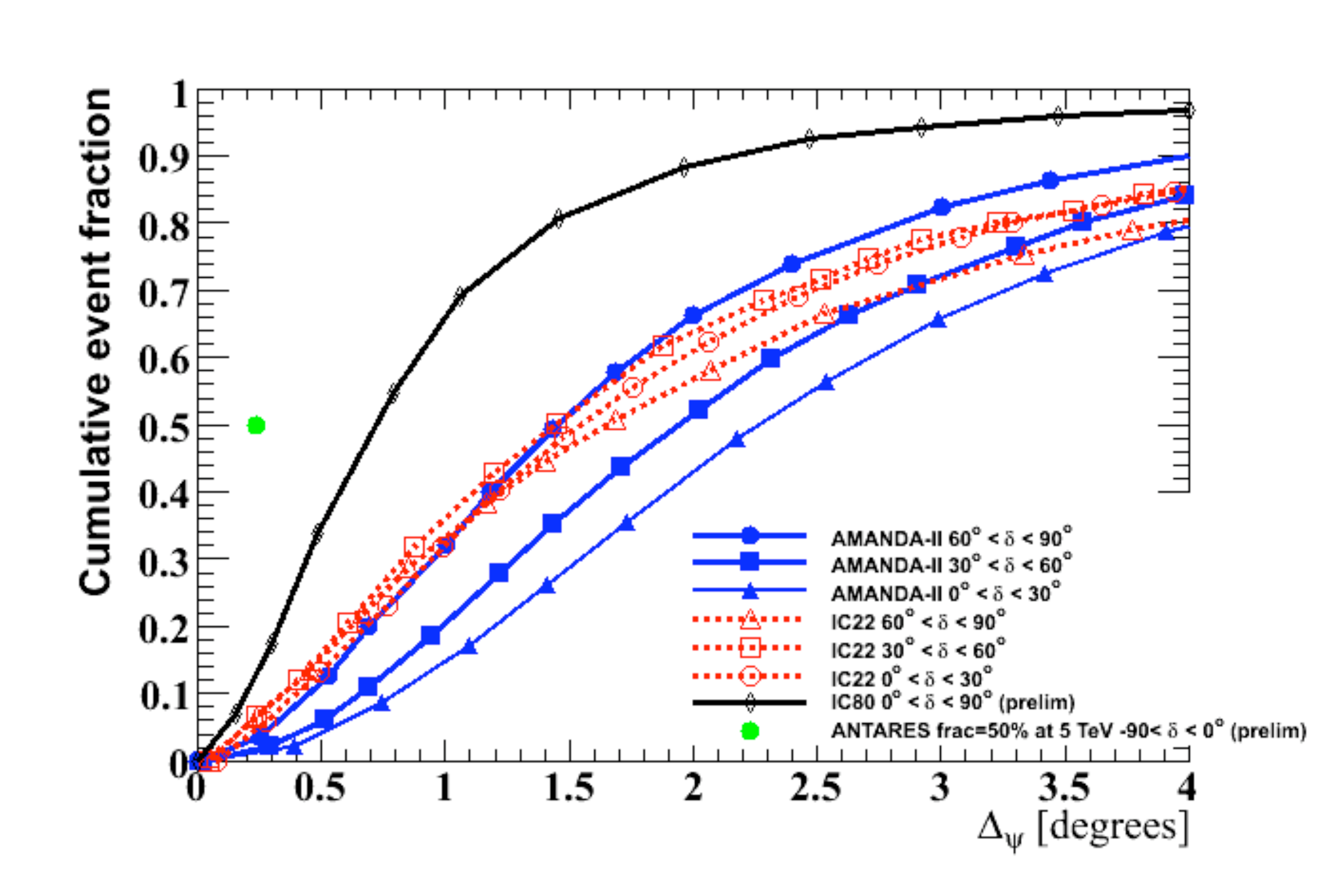}
\caption{PSF (fraction of events reconstructed inside an angular distance from the $\nu$ source direction calculated from full simulations) for an $E^{-2}$ flux of neutrinos for event selections designed for point-like source search studies in IC22 and AMANDA-II \protect\cite{amanda7} in 3 declination regions. For IC80  the expected PSF is integrated over the Northern hemisphere. ANTARES point is explained in the text \protect\cite{transmission_antares}.  This point compares to IC22: $\Delta \Psi = 1.5^{o}$ including 50\% of the events. It should be considered though that for IC22 the analysis has been fully developed, cuts are optimized and agreement data/MC is proved, while for ANTARES this value is based on simulation.
}
\label{fig6}
\end{figure}

\begin{figure}[htb]
\vspace{9pt}
\includegraphics[width=20pc,height=18pc]{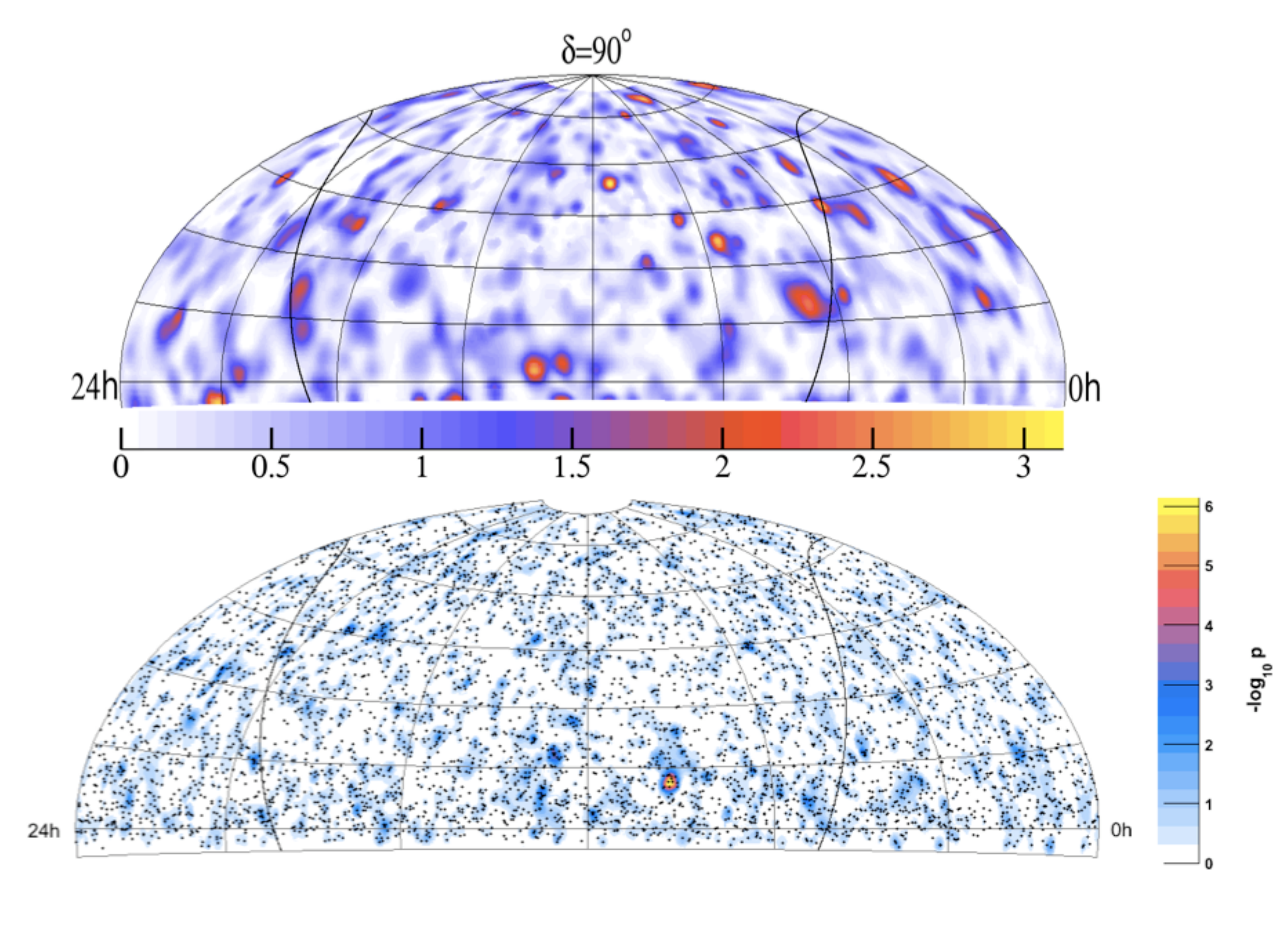}
\caption{Skymaps showing the $log_{10}$ of the p-value (pre-trial) for the 2 data samples of 6595 events for AMANDA-II 1387d (top) and IC22 275.7 d (bottom). Notice the different color code for the pre-trial p-value in the two figures.}
\label{fig7}
\end{figure}

\begin{figure}[htb]
%\vspace{9pt}
\includegraphics[width=18pc,height=18pc]{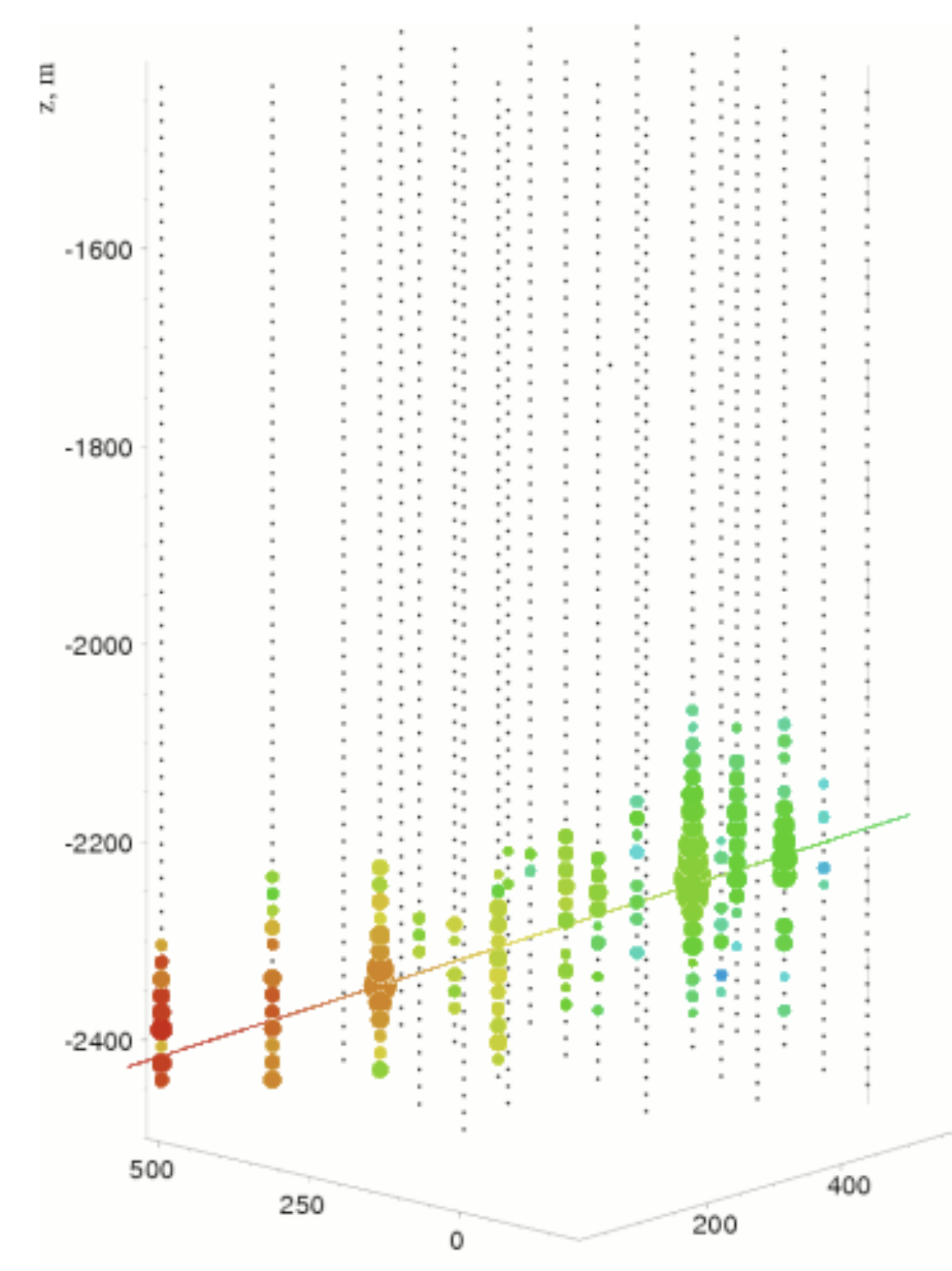}
\caption{One of the highest energy horizontal neutrinos in the hot spot in the IC22 point-source analysis. This event switched on NCh = 145 DOMs on 16 strings deep in the ice where the transparency of the ice allows photons to propagate hundreds of meters. From the MC of atmospheric neutrinos we expect about 2.3 neutrinos with NCh $> 140$ per year from the entire hemisphere and 0.4 from the horizontal declination band between $6^o-16^o$ where we measure 3. Neutrinos producing such high numbers of hit can have energies $\gtrsim 500$ TeV.  }
\label{fig8}
\end{figure}

The current status of various observations for point source searches is shown in Fig.~\ref{fig5}. As can be seen the sensitivity of growing configurations of IceCube is rapidly pushing the explorations of neutrino fluxes for the Northern hemisphere in the region of below $\frac{dN}{dE} \sim 10^{-11}-10^{-12}$ TeV$^{-1}$ cm$^{-2}$ s$^{-1}$ $E^{-2}$. Various models for neutrino production in galactic sources predict fluxes in this range, for instance the microquasar model applied to LS5039 in 
Ref.~\cite{ls5039} and the one concerning the Milagro hot spots \cite{halzen_milagro}. This is also the order of magnitude one gets considering a supernova at a distance of 1 kpc that transfers 10\% of its energy of the order of $10^{51}$ erg to cosmic rays that interact with molecular clouds with 1 cm$^{-3}$ density \cite{halzen}. In the Southern hemisphere the sensitivity of 5-lines of ANTARES for 140 d is comparable to upper limits of MACRO \cite{macro} for 5.6 yrs and of Super-Kamiokande \cite{sk} for 4.5 yrs. In 1 yr the ANTARES sensitivity for 12 lines should be lower by about an order of magnitude. The 22 string configuration (IC22) for a livetime of 275.7 d has a sensitivity of about a factor of 3 lower in the horizontal region compared to 7 yrs of the AMANDA-II detector \cite{amanda7} corresponding to a livetime of 1387 d. The total data sample collected by AMANDA-II is made of 6595 upward-going muon events. IC22 data sample is composed of 5114 upgoing muon events and we expect 4642 atmospheric neutrinos using the Bartol flux calculation \cite{bartol} in 275.7 d. The agreement between data and MC is well inside the systematic uncertainty of atmospheric neutrino calculations of about 15\% \cite{bartol} given the estimated contamination of misreconstructed atmospheric muons of about 10\% larger in the horizontal region. A consistent part of the contamination is due to coincident cosmic rays producing muons from two different directions that confuse reconstructions. 
Selection criteria for obtaining these samples aim at a good angular resolution. The point spread functions (PSF) for AMANDA-II, IC22 and IC80 are shown in Fig.~\ref{fig6} and compared for what possible in this plot to ANTARES: the point at $0.25^{o}$ is taken from a plot of the median angle between reconstructed muons and parent neutrinos as a function of energy at the peak energy of 5 TeV of the response curve for an $E^{-2}$ flux \cite{transmission_antares}. It is noticeable that for IC22 (and this applies also to IC80) the PSF mildly depends on the declination, while for AMANDA it does since the detector's height is much larger than its width. AMANDA is a little better in the vertical direction than IC22 due to the smaller vertical spacing of optical modules along a string ($\sim 10$ m respect to 17 m in IceCube). The ANTARES better angular resolution of is due to the fact that light is much less scattered in sea water than in ice.
The AMANDA-II and IC22 neutrino sky-maps are shown in Fig.~\ref{fig7}. A likelihood method was applied to look for excesses of high energy neutrino events clustered around any direction in the sky or around a catalogue of candidate sources on top of the atmospheric neutrino background. For this analysis event times are scrambled in declination bands in order to reproduce many background-only `equivalent experiments'. A hot spot in the all-sky search corresponding to a 1.3\% post trial probability (p-value) to be a fluctuation of the background is found. Though this p-value is not significant enough to claim any evidence of a neutrino source, enough data have already been collected with IC40 to verify or exclude this as a possible signal. This analysis also uses the energy information based on the fact that the astrophysical neutrinos are expected to have a much harder spectrum than atmospheric neutrino ones. In Fig.~\ref{fig8} we show one of the highest energy events that contributed most to the p-value of the hot spot region.

\section{CONCLUSIONS}
This is a great time for Neutrino Astronomy since new experiments are deriving exciting results and since we are close to the reach of expected neutrino fluxes correlated to gamma astronomy observations.


\begin{thebibliography}{19}
\bibitem{GZK} J. Beltz for HiReS Coll. and C. Bonifazi for the Pierre Auger Coll., these proceedings
\bibitem{PAOanistropies} J. Abraham {\it et al.}, Science 318 (2007) 938.
\bibitem{hires_anisotropy} R. Abbasi {\sl et al.}, Astropart. Phys. 30 (2008) 175.
\bibitem{WB} E. Waxman and J.N. Bahcall, Phys. Rev. D59 (1999) 023002; Phys. Rev. 78 (1997) 2292.
\bibitem{MPR} K. Mannheim, R.J. Protheroe and J.P. Rachen, Phys. Rev. D63 (2001) 023003.
\bibitem{ahlers} M. Ahlers {\sl et al.}, Phys. Rev. D72 (2005) 023001.
\bibitem{sk_diffuse} M.E.C. Swanson {\it et al.}, Astrop. J. 652 (2006) 206.
\bibitem{macro_diffuse} M. Ambrosio {\sl et al.}, Astrop. Phys. 19 (2003) 1.
\bibitem{amanda_diffuse} A. Achterberg {\sl et al.}, Phys. Rev. D76 (2007) 042008.
\bibitem{antares_diffuse} A. Romeyer, R. Bruijn and J. de D. Zornoza for the ANTARES Coll., proc. of ICRC2003, hep-ex/0308074.
\bibitem{performance} J.~Ahrens {\it et al.}, Astrop. Phys. 20 (2004) 507.
\bibitem{amanda_atmnu} K. M\"unich, J. L\"unemann for the IceCube Coll., in proc. of ICRC2007, arXiv:0711.0353.
\bibitem{bartol} G.~Barr {\it et al}, Phys. Rev. D70 (2004) 023006. 
\bibitem{naumov} G. Fiorentini, A.V. Naumov and F.L. Villante, Phy. Lett. B510 (2001) 173.
\bibitem{HKKM} M. Honda {\sl et al.}, Phys. Rev. D 75 (2007) 043006.
\bibitem{martin}  A.D. Martin, M.G. Ryskin and A.M. Stasto, Acta Phys. Polon. B34 (2003) 3273.
\bibitem{anita} S.W. Barwick {\it et al.},  Phys. Rev. Lett. 96 (2006) 171101.
\bibitem{gorham} P~ Gorham, to appear in proc. of Neutrino 2008, Christchurch, New Zealand.
\bibitem{rice} L.~Kravchenko {\it et al.}, Phys. Rev. D73 (2006) 082002.
\bibitem{eheicrc07} A. Ishihara for the IceCube Coll., proc. of ICRC2007, astro-ph/0711.0353.
\bibitem{PAO} J. Abraham {\sl et al.}, Phys. Rev. Lett. 100 (2008) 211101.
\bibitem{hires} R.~Abbasi {\sl et al.}, arXiv:0803.0554.
\bibitem{amandauhe} M.~Ackermann {\it et al}, Astrop. J 675 (2008) 1014.
\bibitem{cascade} M.~Ackermann {\it et al},  Atrop. Phys. 22 (2004) 127.
\bibitem{tauwer} M.~Iori {\sl et al.}, astro-ph/0602108.
\bibitem{allard} D. Allard {\it et al.}, JCAP 0609 (2006) 005.
\bibitem{engel01} R. Engel, D. Seckel and T. Stanev, Phys. Rev. D64 (2001) 093010.
\bibitem{berghaus} P. Berghaus for the IceCube Coll., these proceedings.
\bibitem{margiotta} A. Margiotta for the ANTARES Coll., these proceedings. 
\bibitem{km3} C. Distefano for the KM3NeT Coll., at this conference 
\bibitem{NEMO} C. Distefano for the NEMO Collab., these proceedings.
\bibitem{amanda7} R.~Abbasi {\it et al.}, arXiv:0809.1646, subm.to Phys. Rev. D, (2008).
\bibitem{ls5039} F.A. Aharonian {\it et al.}. J. Phys. Conf. Ser. 39 (2006) 408.
\bibitem{halzen_milagro} F. Halzen {\it et al.}, Phys. Rev. D78 (2008) 063004.
\bibitem{halzen} F. Halzen, arXiv:0809.1874.
\bibitem{macro} M.~Ambrosio {\it et al.}, Astrophys. J.  546 (2001) 1038.
\bibitem{sk} K. Abe {\it et al.}, Astrophys. J. 652 (2006) 198.
\bibitem{taup07} T.~Montaruli for the IceCube Collaboration, J. Phys. Conf. Ser. {\bf 120} (2008) 062009. 
\bibitem{transmission_antares} J.A. Aguilar {\it et al.}, Astropart. Phys. 23 (2005) 131.
\end{thebibliography}
\end{document}